# A global inversion-symmetry-broken phase inside the pseudogap region of YBa$_2$Cu$_3$O$_y$

L. Zhao[1,2], C. A. Belvin[3], R. Liang[4,5], D. A. Bonn[4,5], W. N. Hardy[4,5], N. P. Armitage[6] & D. Hsieh[1,2]

[1]Department of Physics, California Institute of Technology, Pasadena, CA 91125, USA.

[2]Institute for Quantum Information and Matter, California Institute of Technology, Pasadena, CA 91125, USA.

[3]Wellesley College, Wellesley, MA 02481, USA.

[4]Department of Physics and Astronomy, University of British Columbia, Vancouver, British Columbia V6T 1Z1, Canada.

[5]Canadian Institute for Advanced Research, Toronto, Ontario M5G 1Z8, Canada.

[6]Institute of Quantum Matter, Department of Physics and Astronomy, The Johns Hopkins University, Baltimore, Maryland 21218, USA.

**The phase diagram of cuprate high-temperature superconductors features an enigmatic pseudogap region that is characterized by a partial suppression of low energy electronic excitations[1]. Polarized neutron diffraction[2–4], Nernst effect[5], THz polarimetery[6] and ultrasound measurements[7] on $YBa_2Cu_3O_y$ suggest that the pseudogap onset below a temperature $T^*$ coincides with a bona fide thermodynamic phase transition that breaks time-reversal, four-fold rotation and mirror symmetries respectively. However, the full point group above and below $T^*$ has not been resolved and the fate of this transition as $T^*$ approaches the superconducting critical temperature $T_c$ is poorly understood. Here we reveal the point group of $YBa_2Cu_3O_y$ inside its pseudogap and neighboring regions using high sensitivity linear and second harmonic optical anisotropy measurements. We show that spatial inversion and two-fold rotational symmetries are broken below $T^*$ while mirror symmetries perpendicular to the Cu-O plane are absent at all temperatures. This transition occurs over a wide doping range and persists inside the superconducting dome, with no detectable coupling to either charge ordering or superconductivity. These results suggest that the pseudogap region coincides with an odd-parity order that does not arise from a competing Fermi surface instability and exhibits a quantum phase transition inside the superconducting dome.**

The crystal system and point group of a material are encoded in the structure of its second- and higher-rank optical susceptibility tensors[8], which can be determined from the anisotropy of its linear and nonlinear optical responses. The second harmonic (SH) response is particularly sensitive to the presence of global inversion symmetry because unlike the linear electric-dipole susceptibility tensor $\chi_{ij}^{ED}$, which is allowed in all crystal systems, the SH electric-dipole susceptibility tensor $\chi_{ijk}^{ED}$ vanishes in centrosymmetric point groups, leaving typically the far weaker electric-quadrupole term $\chi_{ijkl}^{EQ}$ as the primary bulk radiation source[9,10]. For these reasons, it has been proposed that optical SH generation may be an effective probe of inversion symmetry breaking in the cuprates[11].

To fully resolve the spatial symmetries underlying the pseudogap and its neighboring regions, we performed both linear and SH optical rotational anisotropy (RA) measurements on de-twinned single crystals of YBa$_2$Cu$_3$O$_y$ as a function of oxygen content and temperature. The RA measurements track variations in the intensities of light reflected at the fundamental ($I^\omega$) and SH frequencies ($I^{2\omega}$) of an obliquely incident laser beam, which is resonant with the O *2p* to Cu *3d* charge transfer energy ($\hbar\omega = 1.5$ eV), as the scattering plane is rotated about the *c*-axis (Fig. 1a). The use of a recently developed rotating optical grating based technique[12] made it possible to perform full 360º sweeps of the scattering plane angle ($\varphi$) over large temperature ranges while keeping the incident beam spot (~100 μm) fixed to the same location on the crystal to within a few microns.

We first simulate the RA patterns expected from hole doped YBa$_2$Cu$_3$O$_y$ ($y > 6$) based on its reported orthorhombic *mmm* crystallographic point group[13], which is inversion symmetric and consists of three generators $m_{ac}$, $m_{bc}$ and $m_{ab}$ that denote mirror symmetries across the *ac*, *bc* and *ab* planes. Unlike its tetragonal parent ($y = 6$) compound[14], the point group of hole doped YBa$_2$Cu$_3$O$_y$ only endows the crystal with two-fold ($C_2$) rather than four-fold ($C_4$) rotational symmetry about the *c*-axis because the excess oxygen atoms form chains that run along the *b* axis (Fig. 1a). Figure 1b shows representative linear and SH RA patterns computed in the electric-dipole $I^\omega(\varphi) \propto \left|\hat{e}_i^\omega(\varphi)\chi_{ij}^{ED}\hat{e}_{j,0}^\omega(\varphi)\right|^2 I_0$ and electric-quadrupole $I^{2\omega}(\varphi) \propto \left|\hat{e}_i^{2\omega}(\varphi)\chi_{ijkl}^{EQ}\hat{e}_{j,0}^\omega(\varphi)\partial_k\hat{e}_{l,0}^\omega(\varphi)\right|^2 I_0^2$ approximations respectively, where $I_0$ is the incident beam intensity and both the incident (in) and reflected (out) polarizations $\hat{e}_0$ and $\hat{e}$ are selected to be perpendicular (S$_{in}$–S$_{out}$) to the scattering plane (see Supplementary Section S1 for other polarization geometries). The orthorhombic crystal system is identifiable by an ovular $I^\omega(\varphi)$ pattern with maxima and minima aligned strictly along the *a* and *b* axes[15] while the *mmm* point group is identifiable by a four-lobed $I^{2\omega}(\varphi)$ pattern that is mirror symmetric about the *a* and *b* axes.

Figure 2 shows linear and SH RA data from YBa$_2$Cu$_3$O$_y$ with hole-doping levels ($p$) of 0.125 ($y = 6.67$; underdoped; $T_c = 65$ K), 0.135 ($y = 6.75$; underdoped; $T_c = 75$ K), 0.165 ($y$

= 6.92; optimal doped; $T_c$ = 92 K) and 0.190 ($y$ = 7.0; overdoped; $T_c$ = 86 K) measured at room temperature ($T > T^*$) in $S_{in}$–$S_{out}$ geometry (see Supplementary Section S2 for all other geometries). In the linear RA data, we see that the anisotropy ($\chi_{xx}^{ED} - \chi_{yy}^{ED}$) becomes more pronounced with hole doping as expected due to the filling of Cu-O chains. However, in contrast to Fig. 1b, the intensity maxima and minima are rotated away from the *a* and *b* axes, indicating an absence of $m_{ac}$ and $m_{bc}$ symmetries consistent with a monoclinic distortion. In fact, by using the structure of $\chi_{ij}^{ED}$ for a monoclinic crystal system in the expression for $I^\omega(\varphi)$, excellent fits to the data are obtained (Fig. 2). In the SH RA data, we also find clear violations of the *mmm* point group symmetries at all doping levels. In particular, the alternation in lobe magnitude as a function of $\varphi$ and the rotation of the lobe bisectors away from the *a* and *b* axes indicate an absence of $m_{ac}$ and $m_{bc}$ symmetries consistent with the linear RA data. We find excellent agreement of the data with an electric-quadrupole induced SH response from the centrosymmetric 2/*m* monoclinic point group (Fig. 2), which consists of two generators 2 and *m* that denote $C_2$ and $m_{ab}$ symmetries. On the contrary, fits to electric-dipole induced SH from the non-centrosymmetric 2 and *m* monoclinic point groups, as well as to magnetic-dipole-induced[11] or surface electric-dipole-induced SH from the 2/*m* point group, do not adequately describe the data (see Supplementary Section S3).

The degree of monoclinicity, which can be qualitatively tracked via the angular deviation of the intensity maximum (lobe bisector) in the linear (SH) RA data away from the *a*-axis, decreases monotonically between $y$ = 6.67 and $y$ = 7 as shown in Fig. 2. This suggests that the absence of $m_{ac}$ and $m_{bc}$ symmetries at all measured temperatures above $T^*$ (see Supplementary Section S4) originates from vacancy induced monoclinic distortions of the oxygen sublattice, which are also known to be present in $La_{2-x}Sr_xCuO_4$[16]. This interpretation is further supported by recent SH RA measurements performed near the charge transfer resonance of layered perovskite iridates[9,10], which revealed subtle oxygen sublattice distortions that are difficult to resolve using diffraction based probes.

Having established the point group of hole-doped YBa$_2$Cu$_3$O$_y$ above $T^*$, we proceed to search for changes in symmetry across the strange metal to pseudogap boundary. Previous infrared conductivity experiments[17] have shown that optical transition rates at frequencies well above the pseudogap energy scale ($\hbar\omega \gtrsim 0.3$ eV) do not exhibit any measurable temperature dependence across $T^*$. As a corollary, any temperature dependent change in magnitude of $\chi_{ij}^{ED}$ or $\chi_{ijkl}^{EQ}$ at the frequencies used in this study ($\hbar\omega = 1.5$ eV and $2\hbar\omega = 3$ eV) should be correspondingly weak. Figures 3a-d show the temperature dependences of both the linear and SH intensities measured at a fixed value of $\varphi$ in S$_{in}$–S$_{out}$ geometry. For all four doping levels studied in Fig. 2, we observe no change in the linear response as a function of temperature as expected. Surprisingly however, all of the SH responses exhibit a significant order-parameter-like upturn below a doping dependent critical temperature $T_\Omega$. This dichotomy between the linear and SH responses can naturally be reconciled if bulk inversion symmetry is broken below $T_\Omega$, which would turn on a new and stronger source of electric-dipole induced SH radiation on top of the already existing electric-quadrupole contribution.

By plotting the doping dependence of $T_\Omega$ atop the phase diagram of YBa$_2$Cu$_3$O$_y$ (Fig. 3e), we find that the observed onset of inversion symmetry breaking coincides very well with the pseudogap phase boundary $T^*$ defined by spin-polarized neutron diffraction[2–4], Nernst anisotropy[5], THz polarimetry[6] and resonant ultrasound[7] measurements in the optimal and underdoped regions, suggesting a common underlying mechanism. Moreover, we detect an onset of inversion symmetry breaking even inside the superconducting dome in the overdoped region, which may imply a quantum phase transition slightly beyond $p = 0.20$ near where the pseudogap energy scale has also been extrapolated to vanish[18]. Interestingly, the temperature dependence of the SH intensity shows no measurable anomalies upon crossing either the superconducting or the charge density wave ordering[19,20] temperatures (Fig. 3e) and does not rapidly diminish at lower temperatures like the Nernst anisotropy[5], which may be caused by charge density wave ordering[21]. This shows that unlike the superconducting and charge ordered phases, which are competing Fermi surface instabilities[22], the observed inversion symmetry broken phase is independent of and coexistent with both of them.

To determine which symmetries in addition to inversion are removed from the 2/*m* point group below the pseudogap temperature, we performed a comparative study of the SH RA data measured above and below $T_\Omega$. Figure 4a shows RA patterns measured at 295 K and 30 K in $S_{in}$–$S_{out}$ geometry for the optimal doped sample ($T_\Omega \sim$ 110 K), which is representative of all other doping levels studied (see Supplementary Section S5). Below $T_\Omega$ we see that the SH intensity is enhanced in a $\varphi$ dependent manner that preserves the two-fold rotational symmetry of the pattern, which ostensibly implies that $C_2$ is preserved in the crystal. However this interpretation can be ruled out because the inferred above electric-dipole induced SH radiation is strictly forbidden by symmetry in $S_{in}$–$S_{out}$ geometry for any point group (non-magnetic or magnetic) that contains $C_2$ (see Supplementary Section S3). The only alternative interpretation is that the order parameter below $T_\Omega$ breaks $C_2$ symmetry, but is obscured in the RA patterns because of spatial averaging over domains of two degenerate orientations of the order parameter that are related by 180° rotation about the *c*-axis. This would imply a characteristic domain length scale that is much smaller than our laser spot size and consistently explains the absence of any signatures of $C_2$ breaking below $T^*$ in Nernst effect[5] and THz polarimetry[6] data as well as the need for multiple magnetic domains to refine the polarized neutron diffraction data[2–4,23], which are all integrated over even larger areas of the crystal compared to our measurements.

The set of all non-centrosymmetric subgroups of 2/*m* that do not contain $C_2$ consists of the two independent magnetic point groups 2′/*m* and *m*1′ and their associated magnetic and non-magnetic subgroups (see Supplementary Section S6), where the generators 2′, *m* and 1′ denote $C_2$ combined with time-reversal, $m_{ab}$ and the identity operation combined with time-reversal respectively. Using the two-domain ($\alpha$ = 1, 2) averaged expression $I^{2\omega}(\varphi) \propto \sum_{\alpha=1,2}\left|\hat{e}_i^{2\omega}(\varphi)\chi_{ijkl}^{EQ}\hat{e}_{j,0}^{\omega}(\varphi)\partial_k\hat{e}_{l,0}^{\omega}(\varphi) + \hat{e}_i^{2\omega}(\varphi)\chi_{ijk,\alpha}^{ED}\hat{e}_{j,0}^{\omega}(\varphi)\hat{e}_{k,0}^{\omega}(\varphi)\right|^2 I_0^2$, we were able to reproduce all features of the low temperature SH RA data, including the small peaks around the *b*-axis (Fig. 4a), by applying the structure of $\chi_{ijk,\alpha}^{ED}$ for either the 2′/*m* or *m*1′ point group. The decomposition of the fit into its two single domain components is shown in Fig. 4b to explicitly illustrate the loss of $C_2$. An equally good fit to the data can naturally be achieved using any magnetic or non-magnetic subgroup of 2′/m or m1′ because they necessarily allow the same or more non-zero independent $\chi^{ED}_{ijk,\alpha}$ tensor elements. Therefore, while further removal of symmetry elements from 2′/m or m1′ is not necessary to explain the low temperature data, it cannot be completely ruled out.

In conclusion, our results show that the pseudogap region in $YBa_2Cu_3O_y$ is bounded by a line of phase transitions associated with the loss of global inversion and $C_2$ symmetries. Although previous THz polarimetry measurements on hole-doped $YBa_2Cu_3O_y$ thin films reported the onset of a linear dichroic response near $T^*$ that breaks $m_{ac}$ and $m_{bc}$ symmetries[6], we find that these symmetries are already broken in the crystallographic structure above $T^*$ and are thus necessarily absent in any tensor response that turns on below $T^*$. The low symmetry of the point group (likely 2′/m or m1′) underlying the pseudogap region cannot be explained by stripe[24] or nematic[5] type orders alone, which have also been reported to develop below $T^*$. Instead it suggests the presence of an odd-parity magnetic order parameter, which is consistent with theoretical proposals involving a ferroic ordering of current loops circulating within the Cu-O octahedra[11,25–27], local Cu-site magnetic-quadrupoles[28], O-site moments[29] or magneto-electric multipoles generated dynamically through spin-phonon coupling[30]. Regardless of microscopic origin, our results suggest that this order undergoes a quantum phase transition inside the superconducting dome slightly beyond optimal doping, which may be responsible for the enhanced $T_c$ and quasiparticle mass[31] observed in its vicinity. Interestingly, a similar odd-parity magnetic phase has also recently been found in the pseudogap region of a 5d transition metal analogue of the cuprates[32], which hints at a possibly more robust connection between the pseudogap and this unusual form of broken symmetry.

**Acknowledgements**

We thank D. N. Basov, P. Bourges, B. Keimer, S. A. Kivelson, P. A. Lee, J. W. Lynn, J. Orenstein, S. Raghu, B. Ramshaw, C. Varma and N.-C. Yeh for valuable discussions. This work was support by ARO Grant W911NF-13-1-0059. Instrumentation for the RA measurements was partially supported by ARO DURIP Award W911NF-13-1-0293. D.H. acknowledges funding provided by the Institute for Quantum Information and Matter, an NSF Physics Frontiers Center (PHY-1125565) with support of the Gordon and Betty Moore Foundation through Grant GBMF1250. N.P.A. acknowledges support from ARO Grant W911NF-15-1-0560. Work at the University of British Columbia was supported by the Canadian Institute for Advanced Research and the Natural Science and Engineering Research Council.

**Author contributions**

L.Z., D.H. and N.P.A. planned the experiment. L.Z. and C.A.B. performed the RA measurements and N.P.A. determined the crystal alignment. L.Z., D.H. and N.P.A. analysed the data. R.L., D.A.B. and W.N.H. prepared and characterized the samples. L.Z. and D.H. wrote the manuscript.

**Additional information**

Supplementary information accompanies this paper. Correspondence and requests for materials should be addressed to D.H. (dhsieh@caltech.edu).

**Figure 1**

**Layout of optical anisotropy experiment and predicted response of YBa$_2$Cu$_3$O$_y$.**
**a**, The intensity of light reflected at the fundamental ($I^\omega$) and SH ($I^{2\omega}$) frequencies of an obliquely incident beam is measured as a function of the angle ($\varphi$) between the scattering plane and the *a-c* plane of YBa$_2$Cu$_3$O$_y$. The Cu-O chains run along the *b* axis. The incident photon energy (1.5 eV) is resonant with the O *2p* to Cu *3d* charge transfer transition (inset). The polarization of the incident (in) and reflected (out) beams can be selected to lie either parallel (P) or perpendicular (S) to the scattering plane. **b**, Rotational anisotropy of the electric-dipole induced linear response and electric-quadrupole induced SH response simulated for the orthorhombic crystal system and the *mmm* point group respectively in S$_{in}$–S$_{out}$ geometry. The crystallographic *a* and *b* axes are aligned respectively along the laboratory *x* and *y* axes.

**Figure 2**

**Crystal system and point group of YBa$_2$Cu$_3$O$_y$ above *T*\*.** Polar plots of $I^\omega(\varphi)$ and $I^{2\omega}(\varphi)$ measured at *T* = 295 K in S$_{in}$–S$_{out}$ geometry from YBa$_2$Cu$_3$O$_y$ with **a**, y = 6.67, **b**, 6.75, **c**, 6.92 and **d**, 7.0. The $I^\omega(\varphi)$ and $I^{2\omega}(\varphi)$ data sets, which have intensity error bars of approximately ± 1 % and ± 10 % respectively, are each plotted on the same intensity scale with a scaling ratio of $I^\omega(\varphi):I^{2\omega}(\varphi) \approx 1:3 \times 10^{-11}$. The former are fit to the electric-dipole induced linear response of the monoclinic crystal system (red lines) and the latter are fit to the electric-quadrupole induced SH response of the 2/*m* point group (blue lines). The angular deviations of the maxima of $I^\omega(\varphi)$ and lobe bisectors of $I^{2\omega}(\varphi)$ away from the *a* axis are shaded red and blue respectively for clarity.

**Figure 3**

**Inversion symmetry breaking transition across the phase diagram of YBa$_2$Cu$_3$O$_y$.** Temperature dependence of the linear and SH response of YBa$_2$Cu$_3$O$_y$ with **a**, y = 6.67, **b**, 6.75, **c**, 6.92 and **d**, 7.0 normalized to their room temperature values. Data were taken in S$_{in}$–S$_{out}$ geometry at angles $\varphi$ corresponding to the smaller SH lobe maxima at *T* = 295 K (see Fig. 2). The error bars represent the standard deviation of the intensity over 60 independent measurements. Blue curves overlaid on the data are guides to the eye. The inversion symmetry breaking transition temperatures $T_\Omega$ determined from the SH response are marked by the dashed lines. The width of the shaded gray intervals represent the uncertainty in $T_\Omega$. **e**, Temperature versus doping phase diagram of YBa$_2$Cu$_3$O$_y$. The values of $T_\Omega$ (red circles) coincide with the onset of the pseudogap as defined by spin-polarized neutron diffraction[2–4] (blue squares), Nernst effect[5] (open blue squares), THz polarimetry[6] (purple diamonds) and resonant ultrasound[7] (orange circles). The onset of a polar Kerr effect[33] (green diamonds) and short-range charge order as determined by x-ray diffraction[19,20] (green squares) occur at temperatures below *T*\*. Error bars denote the uncertainty in the onset temperatures. The superconducting transition temperatures of the samples used in our work are denoted by gray circles.

**Figure 4**

**Point group symmetry of YBa$_2$Cu$_3$O$_y$ below T*. a**, Polar plot of $I^{2\omega}(\varphi)$ measured at $T$ = 30 K in S$_{in}$–S$_{out}$ geometry from YBa$_2$Cu$_3$O$_{6.92}$ (squares). The intensity error bars are approximately ± 10 %. The blue curve is a best fit to the average of two 180° rotated domains with 2′/m point group symmetry as described in the text. The fit to the $T$ = 295 K data (pink shaded area) reproduced from Fig. 2c is overlaid for comparison. **b**, Decomposition of the fit to the $T$ = 30 K data (shaded blue area) into its individual domain contributions (shaded orange and purple areas).

**Methods**

**Material growth**. YBa$_2$Cu$_3$O$_y$ single crystals were grown in non-reactive BaZrO$_3$ crucibles using a self-flux technique. The CuO-chain oxygen content was set to y = 6.67, 6.75, 6.92 and 7.0 by annealing in a flowing O$_2$:N$_2$ mixture and homogenized by further annealing in a sealed quartz ampoule, together with ceramic at the same oxygen content. The absolute oxygen content (y) is accurate to ± 0.01 based on iodometric titration. The crystals used in our experiments were de-twinned and aligned to high accuracy by x-ray Laue diffraction.

**Optical RA measurements**. The RA measurements were performed using a rotating optical grating based technique[12] with ultrashort (~80 fs) optical pulses produced from a regeneratively amplified Ti:sapphire laser operating at a 10 kHz repetition rate. The angle of incidence was ~30° and the incident fluence was maintained below 1 mJ cm$^{-2}$ to ensure no laser-induced changes to the samples (see Supplementary Section S7). Alignment of the optical axis to the crystallographic $c$-axis was determined to better than 0.1° accuracy as described in Supplementary Section S8. Reflected fundamental ($\lambda$ = 800 nm) and SH light ($\lambda$ = 400 nm) were collected using photodiodes and photomultiplier tubes respectively. Crystals were sealed in a dry environment during transportation and immediately pumped down to pressures $< 5 \times 10^{-6}$ torr for measurements.

**Fitting procedure**. The high-temperature ($T > T_\Omega$) linear optical anisotropy data were fitted to the expression $I^\omega(\varphi) = A|\hat{e}_i^\omega(\varphi)\chi_{ij}^{ED}\hat{e}_{j,0}^\omega(\varphi)|^2 I_0$ and the high- ($T > T_\Omega$) and low-temperature ($T < T_\Omega$) SH anisotropy data were fitted to the expressions $I^{2\omega}(\varphi) = |A\hat{e}_i^{2\omega}(\varphi)\chi_{ijkl}^{EQ}\hat{e}_{j,0}^\omega(\varphi)\partial_k\hat{e}_{l,0}^\omega(\varphi)|^2 I_0^2$ and $I^{2\omega}(\varphi) = (|A|^2/2)\sum_{\alpha=1,2}|\hat{e}_i^{2\omega}(\varphi)\chi_{ijkl}^{EQ}\hat{e}_{j,0}^\omega(\varphi)\partial_k\hat{e}_{l,0}^\omega(\varphi) + \hat{e}_i^{2\omega}(\varphi)\chi_{ijk,\alpha}^{ED}\hat{e}_{j,0}^\omega(\varphi)\hat{e}_{k,0}^\omega(\varphi)|^2 I_0^2$ respectively. Here $A$ is a constant determined by the experimental geometry, $\hat{e}_0$ and $\hat{e}$ are the polarizations of the incoming and outgoing light in the frame of the crystal, $\chi_{ij}^{ED}$ is the linear electric-dipole susceptibility tensor, $\chi_{ijkl}^{EQ}$ is the SH electric-quadrupole susceptibility tensor and $\chi_{ijk,\alpha}^{ED}$ is the SH electric-dipole susceptibility tensor. The $\alpha = 1$ and $\alpha = 2$ versions of the latter are related by 180° rotation about the *c*-axis. The non-zero independent elements of these tensors in the frame of the crystal are determined by applying the monoclinic crystal symmetries to $\chi_{ij}^{ED}$ and the appropriate point group symmetries ($2/m$ for $\chi_{ijkl}^{EQ}$ and either $2'/m$ or $m1'$ for $\chi_{ijk,\alpha}^{ED}$) and degenerate SH permutation symmetries to $\chi_{ijkl}^{EQ}$ and $\chi_{ijk,\alpha}^{ED}$. These operations reduce $\chi_{ij}^{ED}$ to 5 non-zero independent elements (*xx, xz, yy, zx, zz*), $\chi_{ijkl}^{EQ}$ to 28 non-zero independent elements (*xxxx, yyyy, zzzz, xxyy = xyyx, yyxx = yxxy, xyxy, yxyx, xxzz = xzzx, zzxx = zxxz, xzxz, zxzx, yyzz = yzzy, zzyy = zyyz, yzyz, zyzy, xzyy = xyyz, xyzy, yxzy = yyzx, yxyz = yzyx, yyxz = yzxy, zxyy = zyyx, zyxy, zxxx, xxzx, xxxz = xzxx, xzzz, zzxz, zzzx = zxzz*), and $\chi_{ijk,\alpha}^{ED}$ to 10 non-zero independent elements (*xxx, xyx = xxy, xyy, xzz, yxx, yyx = yxy, yyy, yzz, zzx = zxz, zzy = zyz*). Values of $R^2 > 0.95$ were achieved for all fits shown. However the fitted values of the susceptibility tensor elements are not unique. Therefore the fits serve only to address the qualitative question of whether a certain point group can or cannot reproduce the data. They are not intended to convey any quantitative information about the magnitudes of various tensor elements.

**Data availability**. The data that support the plots within this paper and other findings of this study are available from the corresponding author upon reasonable request.

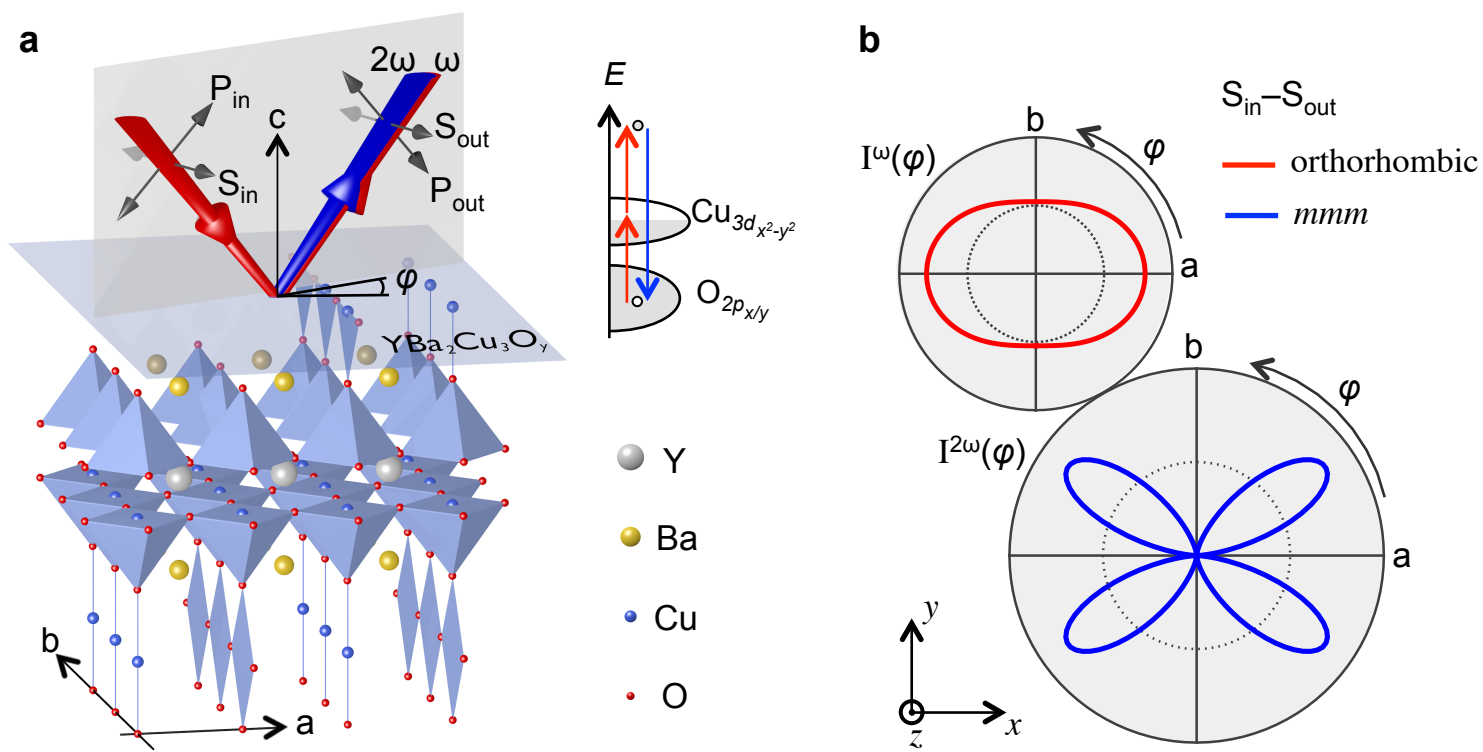

Fig. 1

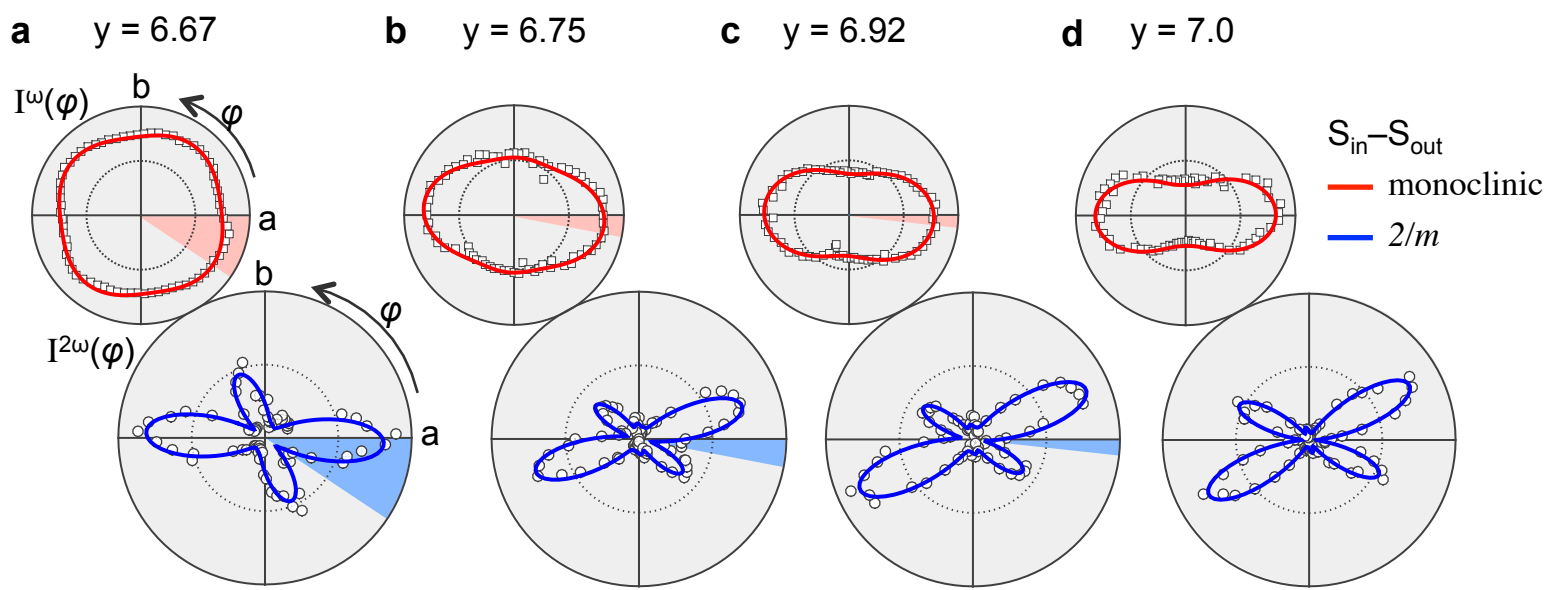

Fig. 2

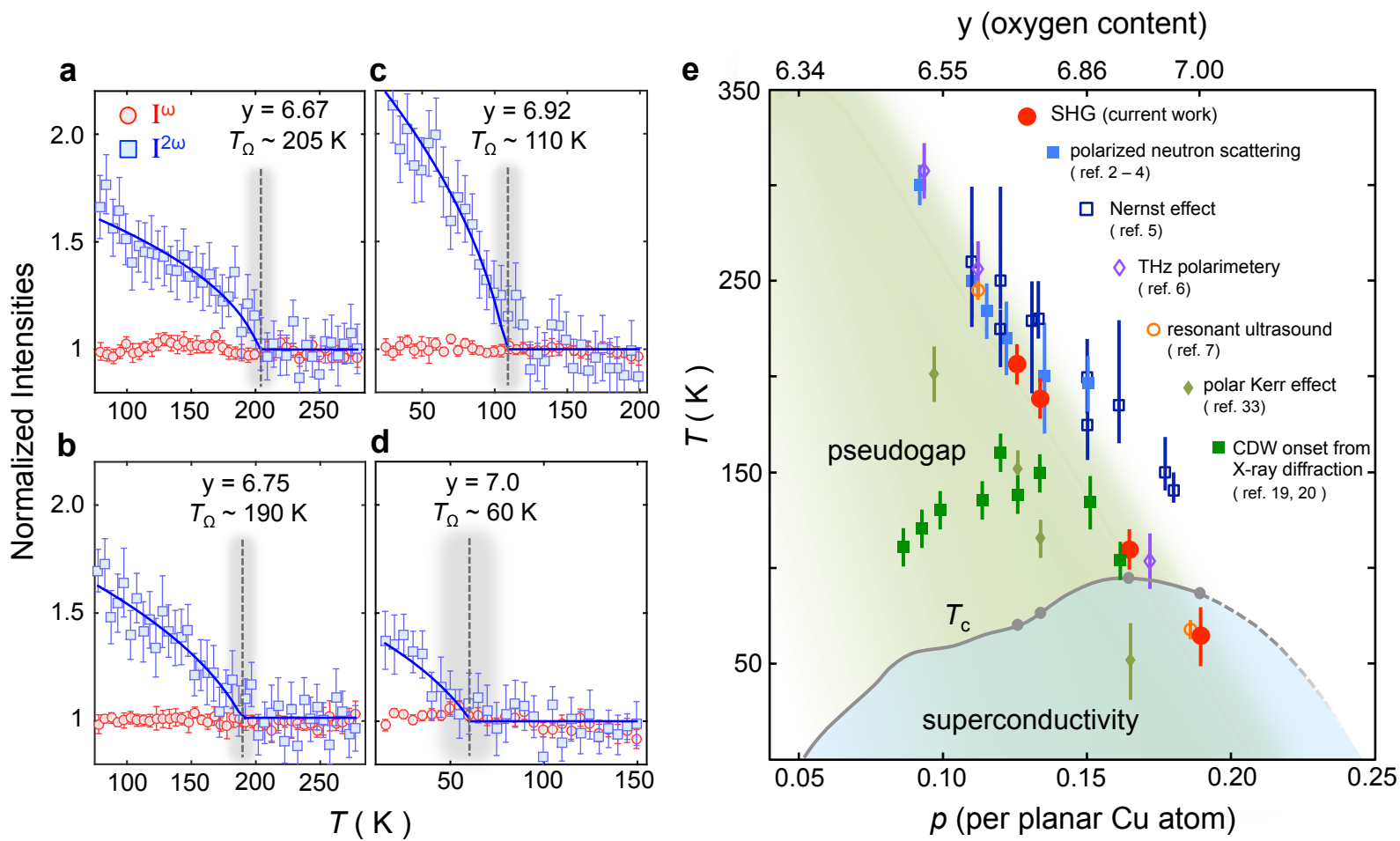

Fig. 3

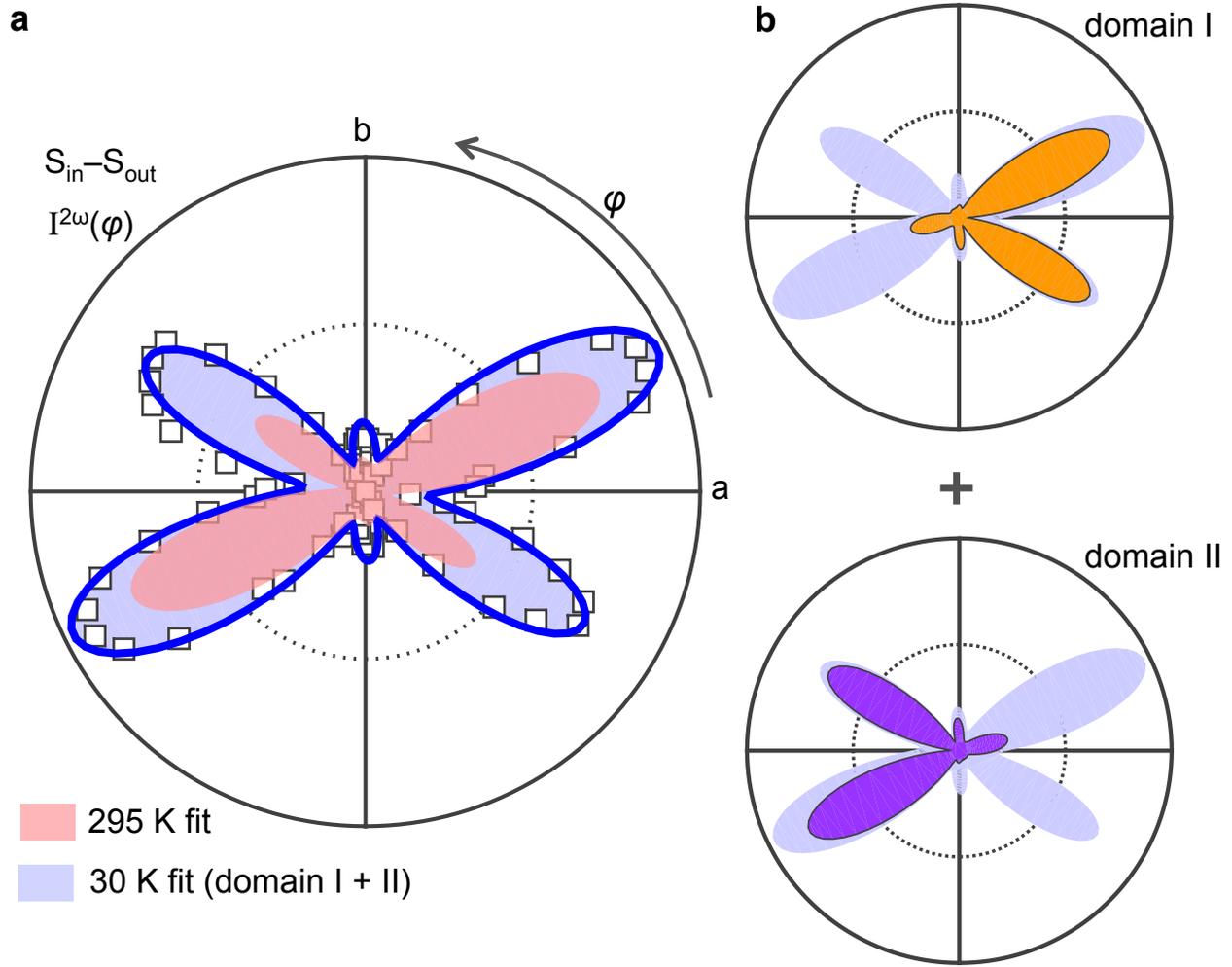

Fig. 4

Supplementary Information for

# A global inversion-symmetry-broken phase inside the pseudogap region of $YBa_2Cu_3O_y$

Contents:

S1. Simulated RA patterns for the orthorhombic $mmm$ point group
S2. RA data for $P_{in}$-$P_{out}$, $S_{in}$-$P_{out}$ and $P_{in}$-$S_{out}$ geometries above $T_\Omega$
S3. Fits of SH RA data above $T_\Omega$ to other monoclinic point groups and radiation sources
S4. Temperature dependence of SH RA data above $T_\Omega$
S5. SH RA patterns above and below $T_\Omega$ for all doping levels
S6. List of subgroups of $2/m$
S7. Exclusion of laser-induced changes to the samples
S8. Exclusion of misalignment as origin of RA patterns

## S1. Simulated RA patterns for the orthorhombic *mmm* point group

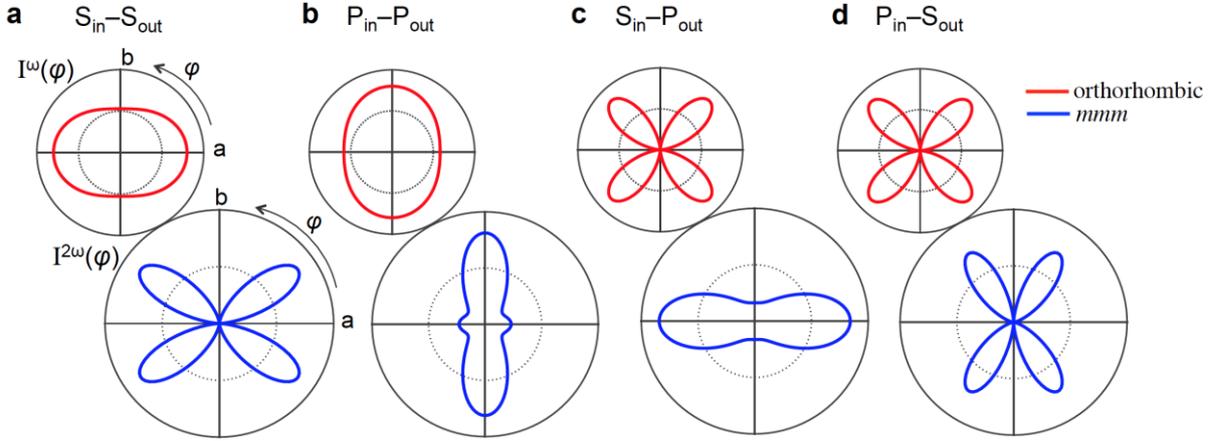

Fig. S1. Simulated linear and SH RA patterns for hole-doped YBa$_2$Cu$_3$O$_y$ using an orthorhombic crystal class and *mmm* point group respectively for all four polarization geometries **a**, S$_{in}$–S$_{out}$ (reproduced from Fig. 1b of the main text), **b**, P$_{in}$–P$_{out}$, **c**, S$_{in}$–P$_{out}$ and **d**, P$_{in}$–S$_{out}$.

We computed linear and SH RA patterns in the electric-dipole and electric-quadrupole approximations for an orthorhombic crystal system and *mmm* point group respectively. The mathematical expressions for $I^\omega(\varphi)$ and $I^{2\omega}(\varphi)$ are given in the Methods section. In the orthorhombic crystal class $\chi_{ij}^{ED}$ has 3 non-zero independent elements (*xx, yy, zz*). In the *mmm* point group and degenerate SH configuration $\chi_{ijkl}^{EQ}$ has 15 non-zero independent elements (*xxxx, xxyy = xyyx, xxzz = xzzx, xyxy, xzxz, yxxy = yyxx, yxyx, yyyy, yyzz = yzzy, yzyz, zxzx, zyzy, zzxx = zxxz, zzyy = zyyz, zzzz*). Figure S1 shows the representative RA patterns computed under all four possible polarization geometries. The P$_{in}$–P$_{out}$, S$_{in}$–P$_{out}$ and P$_{in}$–S$_{out}$ simulations reproduce all of the symmetries of the S$_{in}$–S$_{out}$ data shown in Fig. 1b of the main text (Fig. S1a) as expected.

## S2. RA data for P$_{in}$-P$_{out}$, S$_{in}$-P$_{out}$ and P$_{in}$-S$_{out}$ geometries above $T_\Omega$

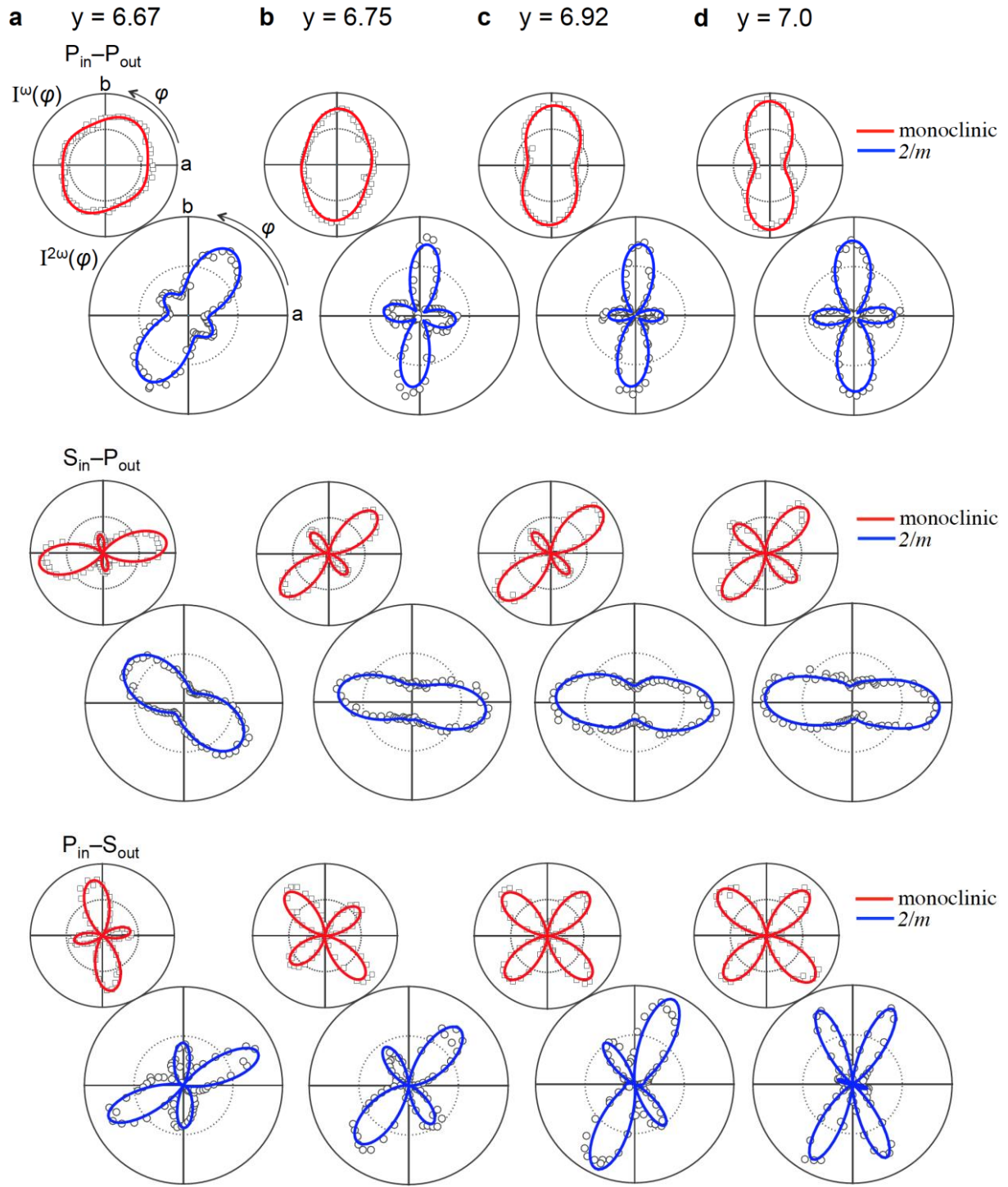

Fig. S2. Linear and SH RA data on hole-doped YBa$_2$Cu$_3$O$_y$ with **a**, y = 6.67, **b**, y = 6.75, **c**, y = 6.92 and **d**, y = 7.0 in P$_{in}$–P$_{out}$ (first row), S$_{in}$–P$_{out}$ (second row), and P$_{in}$–S$_{out}$ (third row) geometries taken at room temperature. Fits to an electric-dipole induced linear response from a monoclinic crystal system (red curves) and to an electric-quadrupole induced SH response from a 2/$m$ point group (blue curves) are overlaid.

Figure S2 shows linear and SH RA data acquired from the four hole-doping levels of YBa$_2$Cu$_3$O$_y$ used in this study under P$_{in}$–P$_{out}$, S$_{in}$–P$_{out}$ and P$_{in}$–S$_{out}$ polarization geometries at room temperature. These data violate the symmetries of the orthorhombic crystal class and *mmm* point group in a manner consistent with the S$_{in}$–S$_{out}$ data shown in Fig. 2 of the main text. Using the same mathematical expressions for $I^{\omega}(\varphi)$ and $I^{2\omega}(\varphi)$ as those used to fit the S$_{in}$–S$_{out}$ data shown in Fig. 2 of the main text (see Methods), excellent fits to the data for all doping levels and polarization geometries were obtained (Fig. S2), thus further confirming our monoclinic 2/*m* point group assignment. Figure S2 also shows that the degree of monoclinicity decreases monotonically between *y* = 6.67 and *y* = 7 for all polarization geometries, which corroborates the S$_{in}$–S$_{out}$ data shown in Fig. 2 of the main text.

## S3. Fits of SH RA data above $T_\Omega$ to other monoclinic point groups and radiation sources

### i. Electric-dipole induced SH from the 2 point group

By applying $C_2$ symmetry and the degenerate SH permutation symmetries alone, $\chi_{ijk}^{ED}$ is reduced to 8 non-zero independent elements (*xxz = xzx, xyz = xzy, yxz = yzx, yyz = yzy, zxx, zxy = zyx, zyy, zzz*), which yields zero response in the $S_{in}$–$S_{out}$ polarization geometry. By extension, any non-centrosymmetric point group that includes $C_2$ as a symmetry element cannot generate an electric-dipole induced SH response in $S_{in}$–$S_{out}$ polarization geometry. Therefore the data shown in Fig. 2 and Fig. 4 of the main text, which have finite $S_{in}$–$S_{out}$ signals, cannot be explained by any non-centrosymmetric point group that contains $C_2$.

### ii. Electric-dipole induced SH from the *m* point group

The *m* point group explicitly violates $C_2$, which is present in all of the room temperature data shown in Fig. 2 of the main text and in Fig. S2. Therefore it is ruled out.

### iii. Magnetic-dipole induced SH from the $2/m$ point group

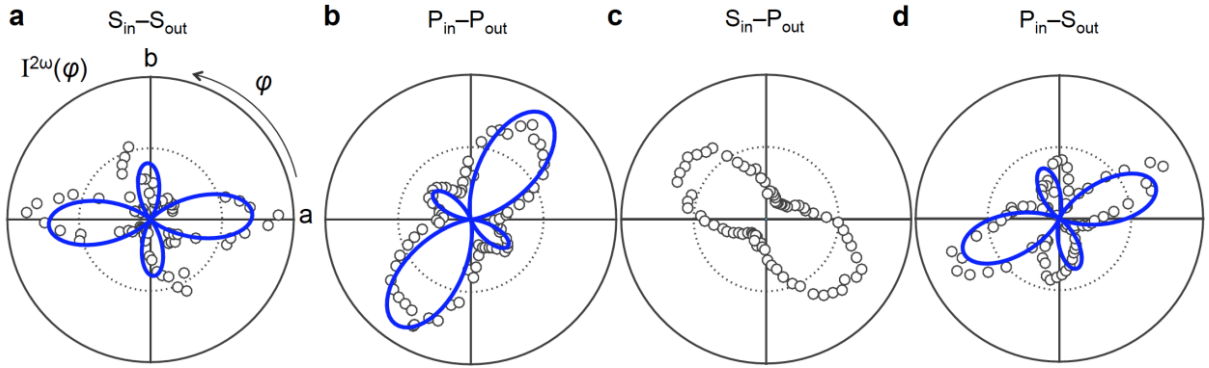

Fig. S3. SH RA data from YBa$_2$Cu$_3$O$_{6.67}$ taken in **a**, $S_{in}$–$S_{out}$, **b**, $P_{in}$–$P_{out}$, **c**, $S_{in}$–$P_{out}$, and **d**, $P_{in}$–$S_{out}$ geometry at room temperature. Blue curves are fits to the bulk magnetic-dipole SH contribution under point group *2/m*.

Magnetic-dipole induced SH is described by the process $M_i^{2\omega} \propto \chi_{ijk}^{MD} E_j^\omega E_k^\omega$ where *M* is the induced magnetization and *E* is the incident electric field. The axial rank-3 magnetic-dipole susceptibility tensor $\chi_{ijk}^{MD}$ is allowed in centrosymmetric crystals. By applying the *2/m* point group symmetries and degenerate SH permutation symmetries, $\chi_{ijk}^{MD}$ is reduced to 8 non-zero independent elements (*xxz = xzx, xyz = xzy, yxz = yzx, yyz = yzy, zxx, zxy = zyx, zyy, zzz*), which yields zero response in the $S_{in}$–$P_{out}$ polarization geometry. This clearly disagrees with the data shown in Fig. S2. Moreover, a best fit to other polarization geometries does not provide good agreement (Fig. S3). Therefore the magnetic-dipole response is ruled out.

### iv.     Surface electric-dipole induced SH from the 2/*m* point group

The surface of any crystal necessarily breaks inversion symmetry and will allow electric-dipole SH generation. However, the (001) surface of a bulk 2/*m* point group contains $C_2$ and therefore yields zero response in $S_{in}$–$S_{out}$ polarization geometry based on the arguments presented in sub-section (i). Therefore the SH response observed above $T_\Omega$ must originate from a source other than the surface electric-dipole contribution.

We note that the additional SH intensity observed below $T_\Omega$ (Figs 3 & 4 of main text) also cannot originate from the surface. First, the temperature at which this symmetry breaking occurs coincides with the known bulk *T\** value for all different doping levels studied. Second, if the enhancement of SH intensity observed below *T\** did in fact originate from the surface, it would imply a corresponding enhancement of the surface electric-dipole radiation efficiency at 1.5 eV incident energy. However the bulk sensitive linear optical response, which is likewise sensitive to electric-dipole transitions at 1.5 eV, shows no detectable change across *T\** (Fig. 3 main text). The only way to reconcile these statements is if a significant change in electric-dipole transition rates at 1.5 eV happened to take place exclusively at the surface across the bulk *T\** value, which to our knowledge has no experimental or theoretical precedent. Therefore we rule out this scenario.

## S4. Temperature dependence of SH RA data above $T_\Omega$

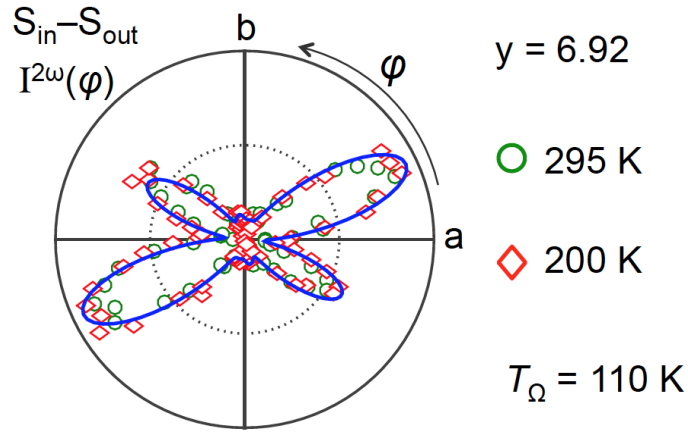

Fig. S4. SH RA data from $YBa_2Cu_3O_{6.92}$ acquired in $S_{in}$–$S_{out}$ polarization geometry at temperatures $T = 295$ K (green circles) and $T = 200$ K (red diamonds), which are both well above $T_\Omega \sim 110$ K. The blue curve is a best fit to the $T = 295$ K data assuming electric-quadrupole induced SH from a $2/m$ point group.

To search for changes in crystallographic structure at temperatures above $T_\Omega$, we performed SH RA measurements as a function of temperature for $T > T_\Omega$. No changes in both intensity and pattern shape were detected above $T_\Omega$ across all four doping levels and polarization geometries. Figure S4 shows representative SH RA patterns from $YBa_2Cu_3O_{6.92}$ in $S_{in}$–$S_{out}$ geometry acquired at $T = 200$ K and $T = 295$ K, which are both well above its measured value of $T_\Omega \sim 110$ K and are nearly identical. This shows that the monoclinic distortion of the lattice sets in above room temperature.

## S5. SH RA patterns above and below $T_\Omega$ for all doping levels

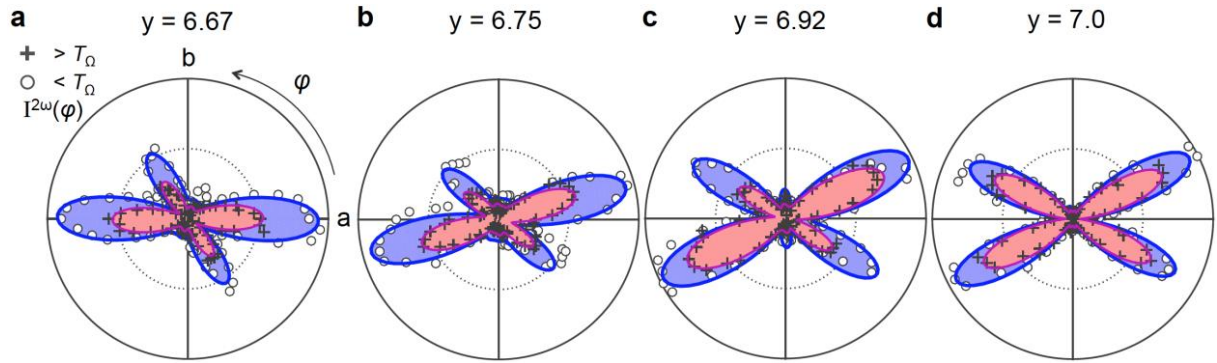

Fig. S5. SH RA data from hole-doped $YBa_2Cu_3O_y$ taken in $S_{in}$–$S_{out}$ polarization geometry both above (crosses) and below (circles) $T_\Omega$. The high temperature data were all acquired at $T = 295$ K while the low temperature data were acquired at **a**, $T = 80$ K for y = 6.67, **b**, $T = 80$ K for y = 6.75, **c**, $T = 30$ K for y = 6.92 and **d**, $T = 15$ K for y = 7.0. Red and blue shaded regions are the best fits to the high and low temperature data respectively using the same models as those described in Fig. 4 of the main text.

Figure S5 shows the SH RA patterns taken both above and below $T_\Omega$ for the complete set of doping levels in $S_{in}$–$S_{out}$ polarization geometry. For all doping levels, the intensity is enhanced below $T_\Omega$ consistent with Fig. 3 of the main text and fit excellently to the same two domain model applied to the data shown in Fig. 4 of the main text.

# S6. List of subgroups of 2/m

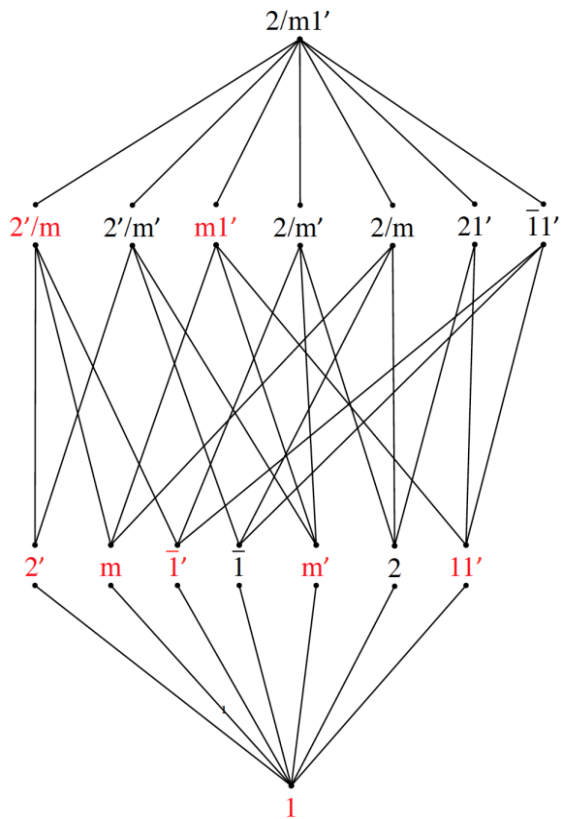

Fig. S6. Tree diagram showing all magnetic and non-magnetic subgroups of the 2/m point group (re-written as 2/m1′). Lines connect parent groups to their subgroups. The non-centrosymmetric subgroups that do not contain $C_2$ are shown in red.

The complete set of magnetic and non-magnetic subgroups of the crystallographic monoclinic point group 2/m is displayed in Fig. S6. The point group 2/m is re-written as 2/m1′ to make the relationship to its magnetic subgroups more obvious. The generators are defined as follows: 2 → 180° rotation about the c-axis, m → reflection about ab-plane, 1 → identity, $\bar{1}$ → spatial inversion, A′ → combination of any spatial operation A with time-reversal. There are a total of 15 subgroups of 2/m1′ (excluding 2/m1′ itself) of which the 8 shown in red are non-centrosymmetric and do not contain $C_2$. The 2′/m and m1′ subgroups are independent (i.e. one is not a subgroup of the other) and contain all of the other 6 red subgroups.

## S7.  Exclusion of laser-induced changes to the samples

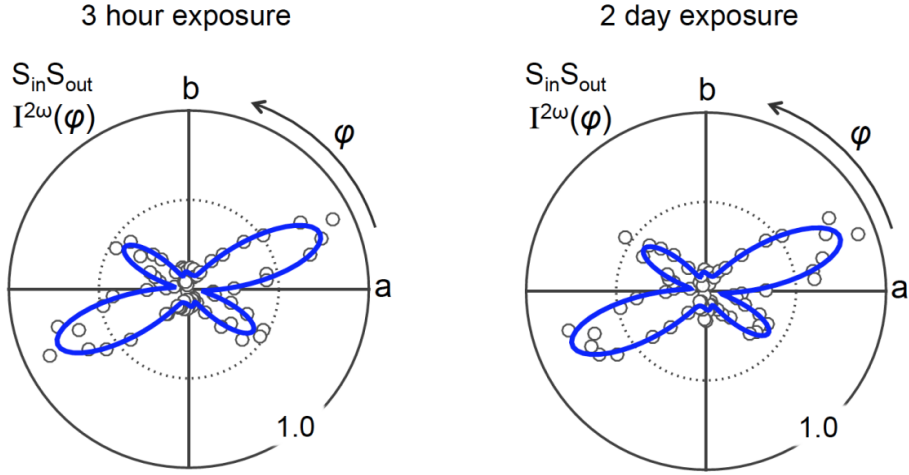

Fig. S7. SH RA data from optimal doped $YBa_2Cu_3O_{6.92}$ taken in $S_{in}$–$S_{out}$ polarization geometry at $T = 295$ K following a 3 hour and 2 day exposure to laser light. Blue curves are fits as described in the main text.

To verify that the laser beam is not inducing any change in the samples, particularly not creating additional oxygen vacancies, we performed SH RA measurements following different laser exposure times. Knowing that the SH RA patterns are highly sensitive to oxygen content (Fig. 2 main text), if the laser were inducing additional oxygen vacancies, one would expect the SH RA patterns to evolve towards those associated with lower oxygen content upon increased laser exposure time. However, we do not observe any obvious change in the SH RA pattern as a function of continuous laser exposure time for any of the samples studied. To give an example, Fig. S7 shows a comparison of two room temperature SH RA patterns measured from the same spot on an optimal doped sample (y = 6.92) after 3 hours versus 2 days of laser exposure plotted on the same intensity scale, which show no measurable difference. [Note: 3 hours is the typical time we require to align the sample]. Moreover, for all doping levels studied, the onset temperatures $T_\Omega$ of the symmetry breaking coincide with the known values of $T^*$ (Fig. 3e main text). Therefore, we rule out any laser-induced changes to the samples.

## S8. Exclusion of misalignment as origin of RA patterns

We can rule out misalignment of any form as the origin of the observed RA patterns for the following independent set of reasons:

1) The flat as-grown faces of the crystals are extremely well aligned perpendicular to the *c*-axis. We know this because atomic force microscopy performed on as-grown crystals show very large micron sized terraces separated by unit-cell tall steps. Therefore any angular misalignment between the surface normal and the *c*-axis would not exceed approximately 1 nm (step height) / 1000 nm (terrace length) = 0.001 radians ≈ 0.06°.

The experimental optical axis is also extremely well aligned perpendicular to the crystal surface. This alignment is performed by ensuring overlap between a collimated incident beam and the retro-reflected beam over a distance of 1 m. Therefore any angular misalignment between the optical axis and the surface normal would not exceed approximately 1 mm (collimated beam diameter) / 1000 mm (beam overlap distance) = 0.001 radians ≈ 0.06°.

The two facts above together demonstrate that the experimental optical axis is extremely well aligned to the crystallographic *c*-axis.

2) The observed deviation from *mmm* symmetry exhibits a systematic dependence on doping (Fig. 2 main text), which is unlikely to arise from random misalignment especially given that the identical alignment procedure is used for all samples.

3) A crystal with *mmm* point group symmetry only has three axes of two-fold rotational symmetry, namely the orthorhombic *a*-, *b*- and *c*-axes. Therefore if the optical axis of the experiment is not along one of these axes, the RA patterns will not exhibit two-fold rotational symmetry. The fact that we do observe two-fold rotational symmetry in the RA patterns (Fig. 2 main text) thus excludes the possibility that what we observe results simply from a small *c*-axis misalignment of an orthorhombic crystal.